\documentstyle[emulateapj5,epsf]{aastex}

\def\lapp{\ifmmode\stackrel{<}{_{\sim}}\else$\stackrel{<}{_{\sim}}$\fi}
\def\gapp{\ifmmode\stackrel{>}{_{\sim}}\else$\stackrel{<}{_{\sim}}$\fi}
\def\dpsr{J0737--3039 system}
\def\psra{PSR\, J0737--3039A}
\def\psrb{PSR\, J0737--3039B}
\def\CXO{{\em Chandra}}
\def\XMM{{\em XMM--Newton}}

\begin{document}


\title{The very soft X-ray spectrum of the Double Pulsar System J0737$-$3039}

\author{A. Possenti\altaffilmark{1}, N. Rea\altaffilmark{2},
  M.~A. McLaughlin\altaffilmark{3,4},
  F. Camilo\altaffilmark{5},  M. Kramer\altaffilmark{6},
  M. Burgay\altaffilmark{1},  B.~C. Joshi\altaffilmark{7}, 
  A.~G. Lyne\altaffilmark{6}}

\altaffiltext{1}{INAF -- Osservatorio Astronomico di Cagliari,
   Loc. Poggio dei Pini, 09012 Capoterra (CA), Italy; possenti@ca.astro.it}
\altaffiltext{2}{University of Amsterdam, Astronomical Institute
   ``Anton Pannekoek'', 1098~SJ, Amsterdam, The Netherlands}
\altaffiltext{3}{Department of Physics, West Virginia University,
  Morgantown, WV 26506, USA}
\altaffiltext{4}{National Radio Astronomy Observatory, 
  Green Bank, WV, 24944, USA}
\altaffiltext{5}{Columbia Astrophysics Laboratory, Columbia University,
  New York, NY 10027, USA}
\altaffiltext{6}{University of Manchester, Jodrell Bank Observatory,
  Macclesfield, Cheshire, SK11 9DL, UK}
\altaffiltext{7}{National Centre for Radio Astrophysics,
  Ganeshkhind, Pune 411007, India}

\begin{abstract}
We present the results of an 80\,ks \CXO\,ACIS-S observation of the
double pulsar system J0737$-$3039. Furthermore, we report on spectral,
spatial and timing analysis of the combined X-ray observations
performed so far for this system. Fitting a total of $\sim 1100$
photons, we show that the X-ray spectrum of the \dpsr\, is very soft,
and not satisfactorily modeled by a simple blackbody or an atmospheric
model. However, it is not possible yet to discriminate between a
predominantly non-thermal and a predominantly thermal origin for the
X-ray emission.

Adopting a simple power-law emission model, the photon index
($\Gamma=3.7\pm0.4,$ 90\% confidence interval) and the implied
conversion efficiency of the rotational energy of \psra\, into X-ray
emission ($4.1\pm0.5\times 10^{-4}$, for a distance to the source of
500 pc) are compatible with the X-ray photons being emitted in the
magnetosphere of \psra. This hypothesis is also supported by the
absence of detectable X-ray orbital modulation (up to $\sim 20\%$,
3$\sigma$) or any X-ray nebular emission and it is in agreement with
the high ($\gapp$ 75\%) X-ray pulsed fraction of \psra.

A two blackbody or a Comptonized blackbody model also reproduce the
data, and the upper limit to the value of the hydrogen column density,
$N_H\la1\times10^{20}$ cm$^{-2},$ is in better agreement (with respect
to the power-law model) with the Galactic $N_H$ in that direction and
at that distance.  For the two blackbody model the implied emission
radii and temperatures are also compatible with those seen in other
recycled pulsars, calling for the bulk of the X-ray photons being
originated from heated regions at the surface of pulsar A. On the
other hand, in the Comptonized blackbody model, the electron
temperature seems to be significantly smaller than in other similar
objects.

\end{abstract}


\section{Introduction}
\label{intro}

The \dpsr\, is a unique celestial object. It comprises a 23 ms pulsar
(\psra, hereafter pulsar A; Burgay et al. 2003\nocite{bdp+03}) and a
2.8 s pulsar (\psrb, hereafter pulsar B; Lyne et
al. 2004\nocite{lbk+04}) revolving every 2.4 hr about their common
center of mass along a somewhat eccentric ($e\sim 0.088$) and highly
inclined ($\sim 89^\circ$; Kramer et al. 2006\nocite{ksm+06}) orbit.

Timing observations in the radio band have allowed the most precise
test of general relativity in strong field to date and promise to
supersede the best Solar-system tests of gravitational theories
(Kramer et al. 2006\nocite{ksm+06}). Various unprecedented phenomena
have also been detected at radio wavelengths, opening the possibility
of studying the so far inaccessible pulsar magnetosphere: e.g. orbital
modulation of the flux density from pulsar B (Lyne et al. 2004),
modulation of the lightcurve of A during the eclipse at the (full and
half) spin period of pulsar B (McLaughlin et al. 2004a), and the
emission of pulsar B being affected by radiation of pulsar A
(McLaughlin et al. 2004b\nocite{mll+04}).

The \dpsr\, is also one of only two double neutron star (DNS) binaries
that have been detected in the X-ray band. This is particularly
interesting, since many different X-ray emission processes may
simultaneously be at work in this system, producing different spectral
signatures.  (1) X-rays could have the quasi-thermal spectrum expected
if the surface(s) of pulsar A and/or pulsar B are heated by a flow of
particles accelerated in the pulsar magnetosphere, depositing their
kinetic energy on the magnetic polar cap(s). Cheng \& Ruderman
(1980)\nocite{cr80} and Arons (1981)\nocite{a81} predicted an almost
uniform temperature for such hot spots, whereas Zavlin \& Pavlov
(1998)\nocite{zp98} suggested that the heated region is larger than
the nominal pulsar polar cap and can be approximated with a central
area at higher temperature surrounded by a larger rim at a lower
temperature.  Alternatively, (2) X-rays could show the power-law
non-thermal spectrum associated with magnetospheric emission.
Applying the outer gap model (see e.g. Cheng \& Zhang
1999\nocite{cz99}), the X-rays are mostly produced by synchrotron
emission in an outer gap, whereas in a polar cap model (e.g. Zhang \&
Harding 2000) X-ray are generated by resonant inverse Compton
scattering off thermal X-ray photons. In both cases the expected
photon index $\Gamma$ of the power-law spectrum is $\lapp 2$, although
observations indicate a somewhat larger range $\Gamma\sim 1-3$ (Zavlin
2006\nocite{z06}; Li et al. 2007\nocite{lll07}). A further possibility
(proposed by Bogdanov et al. 2006 e.g. for the case of
PSR~J0437$-$4715) is that (3) the X-ray emission comprises a thermal
component resulting from polar cap(s) heating and a power-law
component due to inverse Compton scattering of the soft thermal
photons by energetic particles. Processes occurring at the termination
shock may also be observed as in pulsar wind nebulae (PWN; see
Gaensler \& Slane 2006). In particular Lyutikov (2004)\nocite{l04} and
Granot \& M\'esz\'aros (2004)\nocite{gm04} showed that a (4a)
non-thermal X-ray spectrum may originate in the ``collision'' between
pulsar A's wind and pulsar B's magnetosphere. This hyphotesis is
supported by the unusual aforementioned phenomena observed in the
radio band, implying a strong interaction between the energetic flux
of electromagnetic waves and particles released by pulsar A and the
magnetosphere of pulsar B.  Granot \& M\'esz\'aros (2004)\nocite{gm04}
also noticed that (4b) non-thermal X-ray photons with the typical
spectral index of a PWN (1.5--2.1, see e.g. Li et
al. 2007\nocite{lll07}) may be released by the pulsar A's wind just
behind the shock caused by the systemic motion of the binary in the
interstellar medium.

We note that pulsar A is a mildly recycled pulsar with an intermediate
value of inferred surface dipole magnetic field ($B_A\sim 6\times
10^9$ G). The only other similar source detected in the X-ray band
(Kargaltsev et al.  2006) is PSR~B1534+12A having $B\sim
1\times10^{10}$ G, and whose distance ($\sim 1$ kpc; Stairs et
al. 2002\nocite{stt+02}) is about twice as large as that of pulsar A
($d\sim 500$ pc), as inferred from the pulsar dispersion measure and a
model for the Galactic electron density (Cordes \& Lazio
2002\nocite{cl02}; note that a preliminary determination of $d$ from
timing parallax is also consistent with this value, Kramer et
al. 2006\nocite{ksm+06}). Therefore pulsar A is a promising object
for investigating the nature of the X-ray emission in a transition
source between the fully recycled millisecond pulsars (MSPs) with $B$
in the $10^{8}-10^{9}$ G range (see e.g. Zavlin 2006\nocite{z06} for a
review) and the middle-aged or young pulsars having $B\gapp 10^{12}$ G
(see e.g. Li et al. 2007\nocite{lll07} for a listing of those detected
in X-rays).

The results of three \CXO\, and one \XMM\, pointings of the \dpsr\,
have already been published. The first short \CXO\, pointing (see
Table~1) allowed detection of $\sim 80$ photons from the binary
(McLaughlin et al. 2004c\nocite{mcb+04}), whereas the second longer
\XMM\, exposure resulted in $\sim 400$ useful MOS1+MOS2 photons
(Pellizzoni et al. 2004\nocite{pdm+04}). Even the combination of these
two data sets (Campana et al.  2004\nocite{cpb04}; Kargaltsev et al.
2006\nocite{kpg06}) did not provide strong constraints on the spectrum
of the source. Campana et al.\ found that the data were compatible
with both a single blackbody (BB) with effective temperature
$0.20\pm0.02$ keV ($90\%$ confidence level) and emission radius
$75^{+30}_{-9}$ m, or with a single power-law (PL) having photon index
$\Gamma=4.2^{+2.1}_{-1.2}.$ They also found that the combination of a
BB plus a PL with $\Gamma=2$ appeared statistically acceptable,
although the additional PL component was not required by the data.

The \XMM\, pointing allowed the first useful timing analysis, showing
that no variation in the X-ray flux was visible along the orbit, up to
a maximum degree of modulation of $\sim 40\%$ ($99\%$ confidence
level) assuming a sinusoidal shape for the light-curve (Pellizzoni et
al. 2004\nocite{pdm+04}).  The second and third \CXO\, observations
used the high temporal resolution of the HRC-S detector, resulting in
the discovery that the bulk of the X-ray emission from the \dpsr\, is
pulsed at the rotational period of pulsar A (Chatterjee et
al. 2007\nocite{cgm+07}). The X-ray profile is double-peaked with
rapidly rising narrow peaks and a high pulsed fraction of
$74^{+18}_{-14}$ \%. Folding the photons in orbital phase, Chatterjee
et al. (2007)\nocite{cgm+07} likewise did not find any evidence for
X-ray flux variability along the orbit, nor any modulation of the
X--emission at the spin rate of pulsar B\footnote{During the
refereeing process of this paper, the results of an additional \XMM\,
pointing have been presented by Pellizzoni et al. (2008)\nocite{p+08}:
the timing capability of the pn detector and the long duration of the
integration allowed them to detect pulsed X--ray emission also from pulsar
B in part of the orbit}.

In this paper we present (\S\ref{newobs}) a new 80\,ks
\CXO\, ACIS-S observation of the \dpsr, performed on 2006 June,
which resulted in $\sim 500$ additional photons. We then report on
spectral, spatial (\S\ref{spect}) and timing
(\S\ref{timing}) analysis carried out by combining this new dataset
with all previous suitable X-ray observations of this source
(\S\ref{allobs}). The results are discussed in \S\ref{discuss}.

\section{Observations and data reduction}

\subsection{New Chandra data}
\label{newobs}

The \dpsr\, was observed on 2006 June 6--7 (ObsID~5501), for 80\,ks
with the \CXO\, Advanced CCD Imaging Spectrometer (ACIS). The ACIS
CCDs S1, S2, S3, S4, I2 and I3 were on during the observation. The
back-illuminated ACIS-S3 CCD was positioned on the nominal target
position and the source was observed in Timed Exposure (TE) and in
VFAINT mode with a time resolution of 3.241\,s.  Standard processing
of the data was performed by the \CXO\, X-ray Center to Level 1 and
Level 2 (processing software DS 7.6.7.2), while we used CIAO (version
3.4) to complete the data reduction.  All the standard data cleaning
and astrometry correction
procedures{\footnote{http://cxc.harvard.edu/ciao/guides/acis\_data.html}}
have been applied to the data. The source was clearly detected at sky
coordinates (J2000.0) $\mbox{R.A.} = 07^{\rm h}37^{\rm m}51.24^{\rm
s}$ and $\mbox{Decl.} = -30^\circ39^{\prime}40.66^{\prime\prime}$.  We
performed a boresight correction matching 4 X-ray sources lying close
to the \dpsr\, with the 2MASS catalogue, deriving a final positional
error on these coordinates of 0.4$^{\prime\prime}$ at 99\% confidence
level. This position is compatible with the more accurate positions
derived from radio timing (Kramer et al.~2006\nocite{ksm+06}) and
interferometry (Chatterjee, Goss \& Brisken 2005\nocite{cgb05}), and
with that determined from a previous \CXO\, observation (McLaughlin et
al.~2004c\nocite{mcb+04}).

For the timing analysis, we extracted the source events from a
circular region of 2$\arcsec$ radius centered on the source
coordinates (this ensures enclosure of more than 90\% of the source
photons), and we used a 5$\arcsec$ extraction radius for the spectral
analysis. Background events have been acquired from regions of similar
areas, and located on the same S3 CCD, but chosen as far as possible
from the source. The choice of two different extraction radii was
driven by the plan to use the {\it H} statistic in the orbital
modulation search (see \S\ref{timing}), and therefore requiring a
negligible number of background counts in the source event file. For
the spectral analysis this is not an issue, and we have chosen a wider
extraction radius to collect more counts in order to better constrain
the background spectrum, and safely subtract it from the source (see
\S\ref{analysis} and \S\ref{spect}). For both timing and spectral
analyses we used all photons in the 0.3--8\,keV energy range.  The
resulting source background-subtracted count rate is
$6.3\pm0.3\times10^{-3}$ counts s$^{-1}$.

\subsection{Previous Chandra and XMM-Newton datasets}
\label{allobs}

Spectral, spatial and timing analyses (see \S\ref{analysis}) have been
performed joining the observation presented in \S\ref{newobs} with all
the suitable datasets available to date (see Table~1).

The \XMM\, observation of 2004 April was processed using SAS version
7.1.0, cleaned for solar and proton flares (see Table~1 for the
resulting exposure times), and employing the most up-to-date
calibration files (CCF release in 2006 November). The PN camera was
observing in Timing mode, the MOS1 in Prime Full Window mode, and the
MOS2 in Small Window mode (0.03\,ms, 2.6\,s and 0.3\,s timing
resolution for PN, MOS1 and MOS2, respectively). For all of our
analyses we used only MOS1 and MOS2 data, since the PN observation was
highly background dominated (because of the Timing mode set-up).  We
applied an extraction radius for events and spectra of $15\arcsec$ for
MOS1 and MOS2 data (this ensures enclosure of more than 90\% of the
source photons). We used for timing and spectral analyses only photons
in the 0.3--2.2\,keV energy range because above 2.2\,keV the source
was highly background dominated.

The \CXO\, ACIS-S observation taken in 2004 January was re-analyzed
using the same procedures and extraction regions as for the new
dataset presented in \S\ref{newobs}.

The two \CXO\, HRC-S observations performed in 2006 were reprocessed
using standard procedures for HRC
analysis{\footnote{http://cxc.harvard.edu/ciao/guides/hrc\_data.html
}}. We first checked the data for the presence of solar flares and
extracted a new observation-specific bad-pixel file. We then ran a
degap correction, and corrected the astrometry for any processing
offset, starting from Level 1 files. Source and background events have
been extracted from two circular regions of $1.2''$ radius each: one
centered at the source position (this ensures enclosure of more than
90\% of the source photons), and the other as far as possible from the
source.

For all these archival observations we found results which are
consistent with those already published for each dataset (McLaughlin
et al. 2004c\nocite{mcb+04}; Pellizzoni et al. 2004\nocite{pdm+04};
Campana et al. 2004\nocite{cpb04}; Kargaltsev et al.
2006\nocite{kpg06}; Chatterjee et
al. 2007\nocite{cgm+07}). Furthermore, within the limited photon
statistics available, the results from all observations were
consistent with each other. Therefore, when it was relevant/useful, we
added some of the observations together, in order to improve the
statistics.

\subsection{Analysis}
\label{analysis}

For the spectral analysis we made use only of the \CXO\, observation
reported in \S\ref{newobs} and of the \XMM\, observation of 2004 April
(see Table~1).  The \CXO\, HRC-S camera does not have spectral
capabilities, and we have chosen not to include the first \CXO\,
ACIS-S observation because the low number of counts would have
prevented us from using the $\chi^2$ statistic as a measure of the
goodness of our spectral modeling. That left us with a total of
1095$\pm$15 photons (corrected for background) for our spectral
analysis. The spectra have been re-binned (before the background
subtraction) to have at least 15 (for \CXO\,) and 25 (for \XMM\,)
counts per spectral bin; hence we ended with a total of 31 spectral
bins for ACIS-S in the 0.3--8 keV energy range, and 10 and 9 spectral
bins for MOS1 and MOS2, respectively, in the 0.3--2.2 keV energy
range. The different counts per bin used for \CXO\, and \XMM\, data
reflects the need of having a similar signal-to-noise (S/N) per bin
for both observations, hence compensating for the higher background of
the \XMM\, observation.

Response matrices were built for each spectrum in the standard
manner\footnote{For ACIS-S see
http://cxc.harvard.edu/ciao/threads/all.html;\\ for MOS see
http://xmm.vilspa.esa.es/sas/7.1.0/documentation/threads/.}.  Finally,
several emission models were fitted to the data using XSPEC versions
11.3 and 12.1 and adopting an interstellar absorption component
modeled by {\tt phabs} using solar abundances from Lodders (2003). We
added a systematic error of 5\% in order to account for
inter-calibration between the different instruments\footnote{ For more
details see
http://xmm.esac.esa.int/external/xmm$\_$sw$\_$cal/calib/cross$\_$cal/index.php
.}. As a further inter-calibration check, we applied all resulting
models (excluding the 5\% systematics) to the \CXO\, data alone, and
found consistent results.

For spatial analysis we only used HRC-S and ACIS-S datasets, because
of the poorer \XMM\, spatial resolution with respect to \CXO. We built
the instrumental Point Spread Function (PSF) for each of the \CXO\,
observations making use of the {\em ChaRT} and the {\em MARX} software
packages\footnote{http://cxc.harvard.edu/chart/threads/marx/}. For all
the PSFs, we used the source spectrum (see \S\ref{spect}) as an input
for the energy distribution of the PSF itself.  We then created an
image for our ACIS-S and HRC-S PSFs, and fitted it to the data,
searching for any disagreement between either the one or the two
dimensional source profiles and the instrumental PSFs.

For timing analysis we used all the datasets listed in Table~1.  This
gives us an intrinsic time resolution in orbital phase of about
$1/2700$ (set by ACIS-S, which has the worst time resolution of the 5
datasets, see caption of Table~1). Only the portions of any
observation covering an integer number of orbits were considered, thus
ensuring uniform coverage of orbital phase for each instrument; the
last operation left us with 1442 photons ($\sim 125$ of which we
estimate are due to the background) out of a total of 1573 ($\sim 170$
due to the background), spread over 25 orbits (see Table~1).  The
times of arrival (TOAs) of the photons were first referenced to the
barycenter of the Solar System, assuming the accurate radio position
from Kramer et al.~(2006)\nocite{ksm+06} and adopting the JPL
planetary ephemeris DE405. Then, we used
TEMPO\footnote{http://www.atnf.csiro.au/research/pulsar/timing/tempo}
and the timing solution tabulated in Kramer et al. (2006)
\nocite{ksm+06} to calculate the orbital phases (with respect both
to the ascending node of \psra\, and to its periastron) associated
both with the list of 1442 barycentric on-source TOAs and with the
list of barycentric TOAs collected from the background area.
Background-subtracted light-curves as a function of orbital phase
were then produced and inspected for the presence of modulation.

\section{Spectral and spatial results}
\label{spect}

\begin{figure}
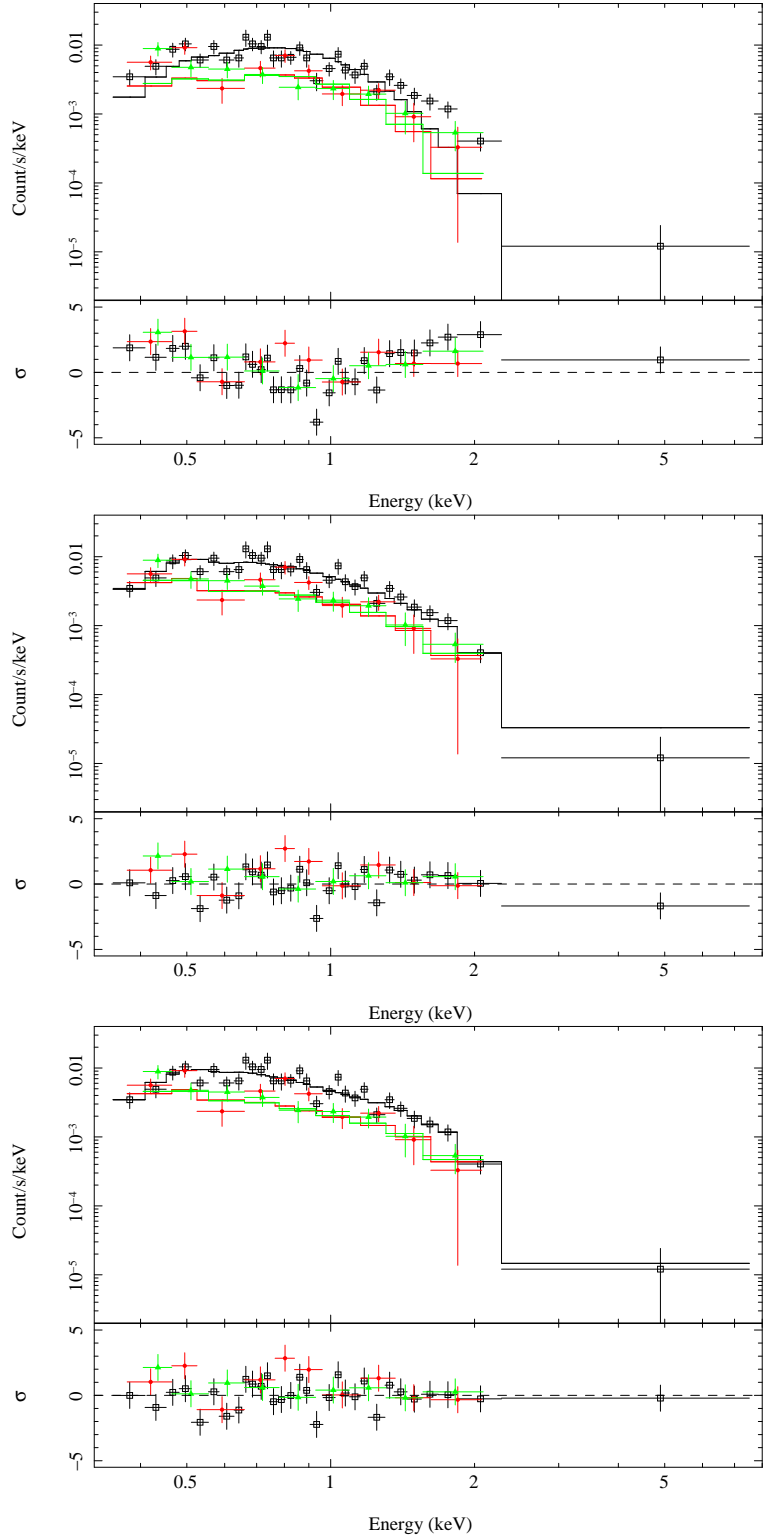

\centering
\vbox{
\includegraphics[angle=270,width=10cm]{f1a.eps}
\includegraphics[angle=270,width=10cm]{f1b.eps}
\includegraphics[angle=270,width=10cm]{f1c.eps}}
\caption{Spectra (top panels) and residulas (bottom panels) of the
spectral modeling of the \dpsr\, obtained from the 2006 June
\CXO\,ACIS-S (black squares) and the 2004 April \XMM\, MOS1 (red filled
circles) and MOS2 (green filled triangles) observations. The data have
been modeled with an absorbed blackbody (top figure), power-law
(middle figure) and two blackbodies (bottom figure; see Table~2 for
details). Solid lines are the fitted models.}
\label{fig:spec}
\end{figure}

We first tried to model the spectrum using single components (see
Table~2). Models consisting of an absorbed blackbody (BB, XSPEC model
{\tt bbody}; see Fig.~\ref{fig:spec}) or an absorbed atmospheric
emission (NSA, XSPEC model {\tt nsa}) are not compatible with the
data. The first has $\chi^2_\nu=2.3$ (see Table~2; $\chi^2_\nu$ is the
reduced $\chi^2$) and a null hypothesis probability (n.h.p.) of
$4\times10^{-6}$.  The second has $\chi^2_\nu=1.8$, and a n.h.p. of
$2\times10^{-3}$.

\begin{figure}
\centering
\includegraphics[angle=270,width=12cm]{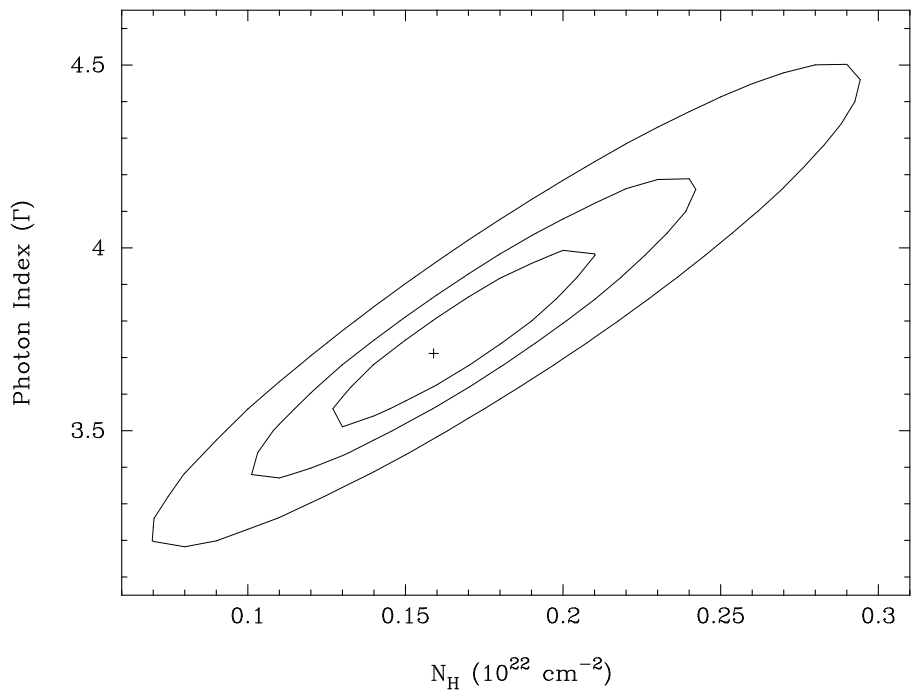}
\caption{Contour plot of N$_H$ vs $\Gamma$ for the absorbed power-law
spectral model (see Table~2). The cross represents the best
fit values, and the ellipses report (from smallest to largest)
the 68\%, 90\% and 99\% confidence level contours on the parameters.}
\label{cntrpl}
\end{figure}

On the other hand, both an absorbed power-law (PL, XSPEC model {\tt
powerlaw}: $\chi^2_\nu=1.25$, n.h.p. of $9\times10^{-2}$) and an
absorbed thermal bremsstrahlung model (BSS, XSPEC model {\tt BREMSS}:
$\chi^2_\nu=1.28$, n.h.p. of $8\times10^{-2}$) can satisfactorily
reproduce our data. The PL model (see Fig.~\ref{fig:spec}) has a large
photon index $\Gamma=3.7\pm0.4$ (here and everywhere in the paper, we
use 90\% confidence intervals, i.e.  $\Delta\chi^2=2.71$) and an
equivalent hydrogen column density
$N_{H}=(1.6\pm0.6)\times10^{21}$cm$^{-2}.$ Figure~\ref{cntrpl} shows a
contour plot of these parameters.  For a distance to the \dpsr\, of
500 pc, the unabsorbed flux (see Table~2) translates into a luminosity
of $(2.4\pm 0.3)\times 10^{30}$ ergs s$^{-1}$ in the 0.3--8 keV range.

The BSS model requires a much lower $N_{H}$ with respect to the PL
model, with an equivalent temperature of $kT = 0.6\pm0.1$ keV and a
0.3--8 keV luminosity of $(0.9\pm0.3)\times 10^{30}$ ergs s$^{-1}.$
However, it is possible to show that\footnote{Given the best fit
values of $kT$ and luminosity, we can follow Grindlay et
al. (2002\nocite{gch+02}) to estimate $\epsilon\sim
\bar{n}^2R^3\sim 7\times 10^{52}$ cm$^{-3}$ (where $R$ is the typical
size of the BSS emitting nebula and $\bar{n}$ is the mean plasma
density in that volume, assuming homogeneity and total charge
neutrality).  Since a large fraction of the X-ray emission from the
system is pulsed (Chatterjee et al. 2007\nocite{cgm+07}) at the spin
period of A, $P_A\sim22.7$ ms, the size of the
emitting region should be $R\lapp cP_A.$ Combining these two constraints, we
can estimate the contribution DM$_{\rm neb}$ of the X-ray BSS emitting
nebula to the dispersion measure; it turns out DM$_{\rm
neb}\sim\bar{n}R\sim \epsilon^{1/2}R^{-1/2}\gapp 10^{22}$
cm$^{-2}.$ Even assuming a negligible contribution due to the plasma
in the intervening interstellar medium (ISM) along the line-of-sight,
the total dispersion measure DM$_{\rm ISM}$+DM$_{\rm neb}$ can hardly
be reconciled with the observed value DM$_{\rm obs}$ for the system.}
these best fit parameters for the emitting nebula would imply
a dispersion measure for the system, $\gapp 10^{22}$ cm$^{-2},$ 
two orders of magnitude larger than the observed value
DM$_{\rm obs}=1.5\times10^{20}$ cm$^{-2}.$  

Given their possible physical relevance, we have also explored the
blackbody plus power-law (BB+PL), the neutron star atmosphere plus
power-law (NSA+PL), the double blackbody (BB+BB) and the Comptonized
blackbody (compBB) models. The combination of a BB (or a NSA) with a
PL does not significantly vary the single PL parameters or improve the
fit, despite the additional free parameters ($\chi^2_\nu\sim1.28$,
n.h.p. of $7\times10^{-2}$). In particular, for all the statistically
acceptable fits, the additional thermal component accounts only for
$\lapp 2\%$ of the total 0.3--8 keV unabsorbed luminosity.  In
Table~2, we report on the BB+PL model showing the maximum contribution
from the blackbody (it has the same BB temperature, $kT=0.18$ keV, of
the best-fit blackbody when fitted alone).  At a distance of 500 pc,
the corresponding BB emission radius is $R_{bb}=18\pm12$ m,
significantly smaller than the nominal polar cap radius of both pulsar
A ($\sim 1$ km) and pulsar B ($\sim 90$ m). Also note that a BB+PL
model with $\Gamma=2$ (still acceptable using only the data taken in
2004; Campana et al.  2004\nocite{cpb04}) is now ruled out
($\chi^2_\nu\sim1.81$, n.h.p.  of $7\times10^{-4}$). On the contrary,
a BB+BB model is compatible with the data (see Fig.~\ref{fig:spec}),
although it does not statistically improve the quality of the fit (see
Table~2): the coolest BB ($kT_1=0.10\pm0.01$ keV) contributes $\sim
60\%$ to the total luminosity with an emission radius
$R_{bb,1}=360\pm150$ m, while the warmest BB ($kT_2=0.30\pm0.05$ keV)
has a tiny $R_{bb,2}=20\pm10$ m.  We note that it has a much lower
$N_{H}$ with respect to the PL model. Following the fitting
prescription by Bogdanov et al. (2006\nocite{bgr06}) for
PSR~J0437$-$4715 (i.e., fixing the value of $N_H$ and assuming a
thermal bath of scattering pairs $e^{\pm}$ at a temperature $\sim
kT_e=150$ keV), a Comptonized blackbody model (compBB, XSPEC model
{\tt compbb}) is completely ruled out ($\chi^2_\nu= 1.95$, n.h.p. of
$8\times10^{-5}$). Allowing $kT_e$ to vary, the $\chi^2_\nu$ improves
significantly (see the best fit parameters in Table~2), although the
fit is still statistically worse than that of the PL or BB+BB models.
The relatively small number of available photons prevent performing a
meaningful fit for even more complicated multi-component models, such
as a two blackbody plus a powerlaw (BB+BB+PL) model or a two
temperature Comptonized model (compBB+compBB).

We note that the $\chi^2_\nu$ resulting from our best fits (those for
the PL model and the BB+BB model) are acceptable, but not very close
to the optimal $\chi^2_\nu=1$ (see Table~1).  As explained in
\S\ref{allobs} and \S\ref{analysis}, we have also fitted the spectra
from \CXO\, and \XMM\, data separately, obtaining compatible
parameters and $\chi^2_\nu$ values in the same range as
above. Therefore inter-calibration should not have a major role in
determining the values of $\chi^2_\nu.$ We have then also searched for
spectral and flux variability using the whole X-ray datasets, and also
dividing them in time--slices (see also \S\ref{timing}), but no such
variability has been detected (however, the limited number of counts
in each time-resolved spectrum makes this non detection not very
constraining). Increasing the number of components in the adopted
model also does not help a lot, as demonstrated e.g. comparing the
$\chi^2_\nu$ of the best fit PL and BB+PL models. Of course, it cannot
be excluded that even more complicated spectral models will finally
improve the values of $\chi^2_\nu.$ However the effect of the
interstellar abundances might also be a promising explanation:
e.g. excluding 3 bins in the spectral fit (one centered at 0.95 keV
and two bracketing the range 0.60-0.75 keV) allows us to reach
$\chi^2_\nu=1.02$ even for the very simple PL model (leaving basically
unchanged the best fit parameters). This improvement may be due to the
presence of a few edges in the photoelectric absorption that are not
properly modeled when adopting solar abundances. We tried to leave the
abundances of the single elements free in {\tt vphabs}, but the low
available counts (compared with the increased number of fitted
parameters) does not make this modeling statistically significant. So,
we think that a much better photon statistics are needed for assessing
if the not optimal value of the best fit $\chi^2_\nu$ is due to a poor
spectral modeling either of the source or of the matter along the
line-of-sight, or something else.

\begin{figure}
\centering 
\includegraphics[angle=270,width=12cm]{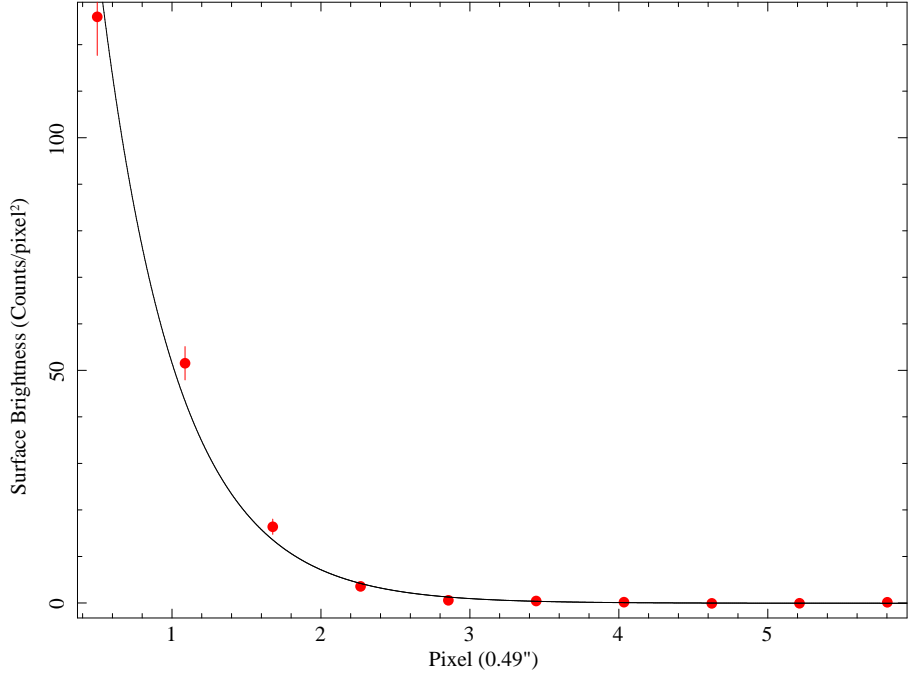}
\caption{\CXO\, ACIS-S one dimensional profile of the \dpsr\, (filled
circles) superimposed with the simulated instrumental PSF (solid line).} 
\label{psf}
\end{figure}

Finally, applying the procedure for the spatial analysis described in
\S\ref{analysis}, we did not detect any diffuse emission neither in the 
\CXO\,HRC datasets (as also reported by Chatterjee et al. (2007) using a 
different analysis method), nor in our new ACIS data . In
Figure~\ref{psf} we report on the comparison between the new ACIS data
of the \dpsr\, with the one dimensional \CXO\, PSF, built as described
in \S\ref{analysis}, as a function of the angular distance from the
source position. Assuming a power-law spectra spanning the range
$\Gamma=2$--4, this translates to a conservative (averaged over an
annulus of 0.5\arcsec--2\arcsec) upper limit on the X-ray luminosity
of a diffuse component of $\sim2\times10^{30}$ ergs s$^{-1}$ (90\%
confidence level, 0.3--8 keV band).

\section{Limits on orbital modulation}
\label{timing}

\begin{figure}
\centering
\includegraphics[angle=0,width=11cm]{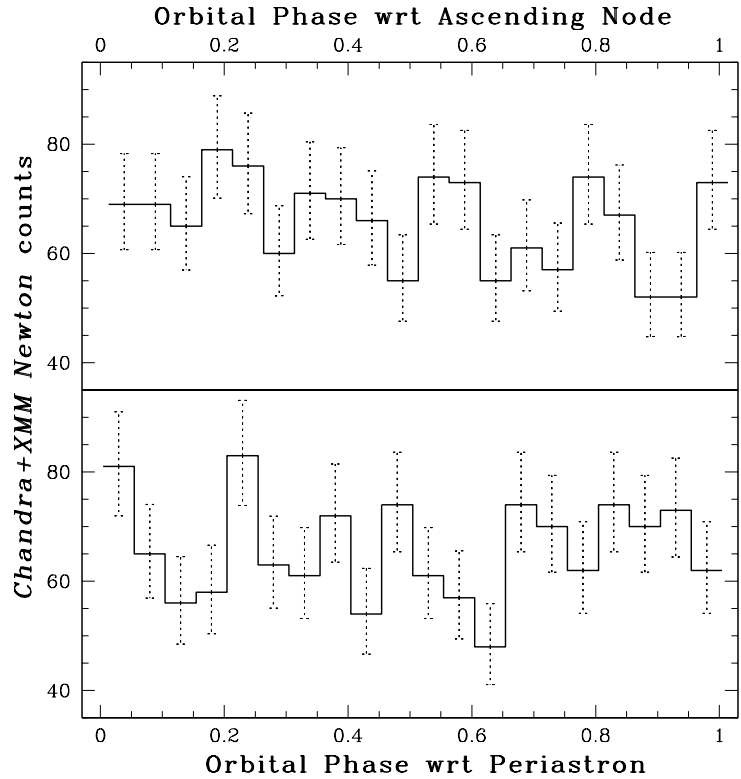}
\caption{Background-subtracted X-ray orbital light-curves for
the \dpsr. The histograms are obtained by folding into 20 bins all the
photons collected so far by \CXO\, and \XMM\, (see Table~1). The
error-bars are plotted following the prescription of Gehrels
(1986). In the lower panel, phase=0.0 is set at the periastron of
\psra\, and the center of the first bin is at phase=0.030, while
in the upper panel phase=0.0 corresponds to the ascending node of
the orbit of \psra\, and the center of the first bin is at
phase=0.039.} \label{lc}
\end{figure}

Figure~\ref{lc} shows the background-subtracted light-curves of the
\dpsr\, obtained by folding all the available X-ray photons
according to the procedure described in \S\ref{analysis}, and
binning into 20 orbital phase bins. Since the periastron of the system
advanced by about 0.11 in orbital phase during the time elapsed
between the first and the last observation, we produced histograms
which are phased both with respect to the periastron of the orbit
and with respect to the ascending node of the orbit of pulsar A. The
first choice (hereafter light-curve LC$_{\rm per}$) aims to reveal
possible modulations in the X-ray curve related to the changing
relative position and orientation of the two neutron stars along
the orbit (e.g. the variation of their distance). The second
option (hereafter light-curve LC$_{\rm asc}$) may better reveal
evidence for modulations related to the orientation of the system
with respect to the line-of-sight.

No statistically significant variations in the X-ray flux were
detected in either of the two folded light-curves. In particular,
applying Pearson $\chi^2$ statistics gives $\chi^2_\nu=1.28$ and
$\chi^2_\nu=1.07$ (with $\nu=19$ degrees of freedom) for LC$_{\rm
per}$ and LC$_{\rm asc}$, respectively, corresponding to probability
of $\sim 17\%$ and $\sim 37\%$ that the histograms are drawn from a
uniform distribution.  Since the results of Pearson statistics depend
on the adopted center of the bins, the values of $\chi^2_\nu$ reported
above are the average over 100 different choices of bin center for
light-curves with 20 bins. We have folded the data with various
combinations of number of bins (from 4 to 32) and number of trial bin
center (in agreement with the typical number of photons in each bin);
all the resulting light-curves are compatible (at $3\sigma$) with a
constant photon flux along the orbit.

In order to overcome the dependency of the results on the binning, we
also adopted the $H$ statistic, which appears to be the best test for
a wide range of physically plausible light-curves (De Jager et
al. 1989\nocite{dsr89}). The value of $H$ is calculated as $H={\rm
Max}(Z^2_m-4m+4),$ where $Z^2_m$ is the family of Rayleigh statistics
and the smoothing parameter $m$ spans the interval $1-20.$
Unfortunately, this test cannot be safely applied to the \XMM\, data,
since the unbinned data are contaminated (at $\sim 25\%$ level) by the
background photons. Limiting the analysis to the 1010 \CXO\, photons
(of which only $\lapp 2\%$ may be due to the background) results in
$H=2.24$ at $m=1$ when the orbital phases are referred to the
periastron and $H=2.38$ at $m=1$ when referred to ascending node.
Hence, the chance probabilities that we are sampling a uniform
distribution are $\sim 41\%$ and $\sim 39\%,$ respectively.

Finally, we have tried to estimate an upper limit to the pulsed
fraction (i.e. the degree of modulation) $p_f=({\rm Max}-{\rm
Min})/({\rm Max}+{\rm Min})$ of the orbital light-curves. In order to
do that, we have used a Monte Carlo code for simulating sinusoidally
modulated light-curves, having on average the same count-rate per bin
of the observed light-curves. For any given value of $p_f,$ about 1000
of these light-curves (with randomly chosen phase) were generated and
then subjected to the Pearson statistic in order to estimate the
probability of their being drawn from an uniform distribution. This
yielded, assuming a sinusoidal modulation of arbitrary phase,
$p_f<18\%$ at $3\sigma,$ a limit more than 2 times smaller than that
of Pellizzoni et al. (2004)\nocite{pdm+04}.

It has been shown (Chatterjee et al. 2007\nocite{cgm+07}) that a large
fraction of the X-ray emission from the system is pulsed at the spin
period of pulsar A. Therefore, our conclusions above mostly constrain
orbital variations of the pulsed emission from pulsar A.  Only HRC-S
data have enough time resolution for selecting photons in the
off-pulse phase of the rotation of pulsar A ($\sim 20\%$ of the pulsar
A's spin period); the analysis of the orbital variations for this
small subset of photons ($\sim 30$) was first performed by Chatterjee
et al. (2007)\nocite{cgm+07}. Repeating their analysis, we likewise
did not find any significant modulation. Given the poor photon
statistics, our derived upper limit to the pulsed fraction of
$p_f\lapp 90\%$ (at $3\sigma$ for a sinusoidal modulation) is not very
constraining.

\section{Discussion}
\label{discuss}

The $\sim500$ additional photons made available by the new \CXO\,
observation allowed us to significantly reduce the uncertainties on
the spectral parameters of the \dpsr and to confirm that it has a very
soft X-ray spectrum. In particular, in contrast to previous attempts
based on a much smaller number of photons (McLaughlin et
al. 2004c\nocite{mcb+04}; Pellizzoni et al. 2004\nocite{pdm+04};
Campana et al. 2004\nocite{cpb04}), our analysis shows that a simple
absorbed blackbody model does not acceptably describe the X-ray
spectrum of the \dpsr\, (see Fig.~\ref{fig:spec}). The same holds true
for a simple thermal emission model (assuming a uniform temperature of
emission) corrected for the radiative transport in the atmosphere of
the neutron star.  Although it is compatible with the data, a thermal
bremsstrahlung spectrum is also excluded on the basis of a comparison
with the observed value of the dispersion measure (see \S\ref{spect}).

However, the $\sim 1100$ available photons do not allow us yet to
discriminate between a predominantly thermal and a predominantly
non-thermal emission. In the following, we discuss these two options
and their ramifications.

\subsection{Thermal emission: heated polar caps}
\label{thermal:bb+bb}

Thermal emission released from relatively small areas having
unequal temperatures is not unexpected in recycled pulsars
(Bogdanov et al. 2006\nocite{bgr06}; Zavlin
2006\nocite{z06}). As reviewed in \S\ref{intro}, it could be
ascribed to the effects of the bombardment of the neutron star
surface by particles accelerated in the magnetosphere of pulsar A, which
is the most energetic of the two pulsars (cooling of the neutron
star interior cannot yield significant X-ray emission for pulsars
older than $\sim 50$ Myr, as appears to be the case for the \dpsr;
Lorimer et al. 2007\nocite{lfs+07}).  The temperatures of the
BB+BB model for the \dpsr\, (see Table~1) are somewhat higher and
the emission radii are smaller than those inferred (Zavlin
2006\nocite{z06}) for the MSPs J0437$-$4715
($kT_1\sim0.04$ keV, $kT_2\sim0.12$ keV, $R_{\rm bb,1}=2.6$ km,
$R_{\rm bb,2}=0.4$ km) and J0030+0451 ($kT_1\sim0.07$ keV,
$kT_2\sim0.18$ keV, $R_{\rm bb,1}=1.4$ km, $R_{\rm bb,2}=0.1$ km)
by applying a model in which the magnetic poles are surrounded by
a weakly magnetized hydrogen atmosphere (Zavlin et al.
1996\nocite{zps96}).  The discrepancy could possibly be
corrected by likewise including the effects of the atmosphere in
the \dpsr\, spectral model. (In fact, a neutron star H-atmosphere
spectrum tends to decrease the best-fit temperature and to
increase the emission radius with respect to a simple blackbody).

The presence of hot spot(s) may be also responsible for the observed
double peaked X-ray light-curve, modulated at the spin period of
pulsar A (Chatterjee et al. 2007\nocite{cgm+07}). A difficulty with
this interpretation is due to the very high observed pulsed fraction
$p_f=74^{+18}_{-14}\%$ ($1\sigma$ error; Chatterjee et
al. 2007\nocite{cgm+07}) of the X-ray light-curve. In fact, the strong
gravitational bending which is experienced by radiation emitted at the
surface of a non magnetized neutron star limits the pulsed fraction
produced by an isotropically emitting hot spot to $\lapp 30\%$
(Psaltis et al. 2000\nocite{pod00}).  Only the presence of strong
surface magnetic fields $B_s\gg10^{10}$ G can significantly increase
the pulsed fraction (Page \& Sarmiento 1996\nocite{ps96}; Geppert et
al. 2006\nocite{gkp06}), perhaps helping to account for sources like
PSR~J1119$-$6127 (a young, high-$B$ pulsar), showing a thermal
spectrum and $p_f\sim 75\%$ (Gonzalez et al 2005\nocite{gkc05}).
However, the value of $B_s$ for PSR~J1119$-$6127 ($4.1\times 10^{13}$
G) is well above that for pulsar A and all the other recycled
pulsars. Thus, anisotropy in photon transport across a weakly
magnetized neutron star atmosphere (e.g. Zavlin et
al. 1996\nocite{zps96}) has been invoked to increase the predicted
pulsed fraction up to the values of 35--50\% seen in the four recycled
MSPs (including the aforementioned PSRs J0437$-$4715 and J0030+0451)
whose X-ray emission is interpreted to be mostly thermal. It is still
debatable if these atmospheric effects can really produce light-curves
approaching as high values of $p_f$ as those seen in pulsar
A\footnote{Note that the pulsed fraction computed for the X-ray
emission from pulsar A is calculated assuming that the whole X-ray
luminosity comes from pulsar A. If other emission mechanisms
contribute to the observed luminosity, the $p_f$ of pulsar A may be
significantly larger.}.  We also note that the X-ray pulse profiles of
the thermally emitting MSPs display more sinusoidal and less spiky
pulses than those shown by pulsar A.

\subsection{Thermal emission: Comptonized spectrum}
\label{thermal:comptbb}

For PSR J0437$-$4715, Bogdanov et al. (2006\nocite{bgr06})
proposed that the predominantly thermal nature of the X-ray emission
(due to heated polar caps) is complemented by the non-thermal emission
due to weak Comptonization of the thermal (blackbody or hydrogen
atmosphere) polar cap emission by energetic electrons/positrons of
small optical depth, presumably in the pulsar magnetosphere and
wind. In the case of the \dpsr, this interpretation (see hypothesis
(3) in \S\ref{intro}) leads to a spectral modeling (see Table~2)
having best fit parameters dissimilar  with respect to
those of PSR J0437$-$4715 ($kT_1\sim0.01$ keV, $kT_2\sim0.25$ keV,
$R_1\sim 0.3$ km, $R_2\sim0.04$ km, and optical depth $\tau_1\sim
0.09$ and $\tau_2\sim 0.06,$ respectively, Bogdanov et
al. 2006\nocite{bgr06}). On a physical ground, this spectral modeling
for the X-ray emission nicely fits with the hypothesis (see Harding \&
Muslimov 2002\nocite{hm02}) that inverse Compton scattering (ICS) is
the main responsible for pair production and particle bombardment of
the polar caps in millisecond pulsars (Bogdanov et
al. 2006\nocite{bgr06}). We note that future deep observations in the
optical band of the \dpsr\, may help discriminating this model with
respect to a pure PL model of magnetospheric origin (see
\S\ref{nonthermal:magn}) and a BB+PL model (see
\S\ref{nonthermal:others}): in fact no non-thermal contribution at
optical wavelengths is expected from a Comptonized spectrum.  On the
other hand, it is not clear if the inclusion of the contribution of a
tail of Comptonized photons to a predominantly thermal spectrum can
alleviate the problem of the high pulsed fraction of the X-ray
emission from pulsar A (see \S\ref{thermal:bb+bb}).

\subsection{Non thermal emission: a magnetospheric origin}
\label{nonthermal:magn}

A non-thermal nature for the \dpsr\, spectrum implies that the bulk of
the X-ray emission ($\gapp 98\%$ in total luminosity) is described by
a power-law with a steep spectral index $\Gamma>3.3$ (at 90\%
confidence level).  As to the origin of this emission, the most
probable hypothesis calls for radiation released by charged particles
accelerated in the magnetosphere of pulsar A. A simple scaling law
$L_x\propto\dot{E}^\beta$ between X-ray luminosity and spin-down
luminosity for recycled pulsars in the Galactic field ($\beta\sim
1.1$; Grindlay et al. 2002\nocite{gch+02}) predicts a 0.5--2.5 keV
luminosity of $\sim 5\times 10^{30}$ ergs s$^{-1}$ for pulsar A.  This
is compatible with the observed 0.5--2.5 keV luminosity of $\sim
10^{30}$ ergs s$^{-1},$ considering the typical scatter (one order of
magnitude) for the above correlation\footnote{Applying the correlation
of Possenti et al. (2002\nocite{pcc+02}; having $\beta\sim 1.4$ in the
2--10 keV band), the predicted luminosity is about 20 times larger
than that observed. The same holds true for the correlation obtained
by Cheng et al. (2006\nocite{ctw06}) using a larger database.  This
large discrepancy may reflect the soft spectrum of pulsar A, whose
luminosity is mostly confined below 2 keV, outside the band selected
by Possenti et al. (2002) and Cheng et al. (2006\nocite{ctw06})}.  The
constancy of the emission along the orbit (\S\ref{timing}) supports
this hypothesis, as does the shape of the X-ray pulse profile
(Chatterjee et al. 2007\nocite{cgm+07}), which shares common features
(i.e. narrow peaks and rapid rise and decline from the maxima) with
the light-curves of the 4 well-studied recycled MSPs having a
predominantly non-thermal spectrum (J0218+4232, B1821$-$24, B1937+21,
and B1957+20; see Zavlin 2006\nocite{z06} for a gallery). Moreover,
the $p_f$ of the pulse profile nicely falls in the range of values
($65-100\%$) observed for the four sources mentioned above.

A problem with the non-thermal emission hypotheses (it applies to
\S\ref{nonthermal:others} as well) is given by the derived value of
$N_H\sim 1.5\times10^{21}$ cm$^{-2}.$ While the value actually matches
that expected for the the dispersion measure (assuming the typical
average of 10 neutral H-atoms for each $e^-$ along the line-of-sight),
the \dpsr\, is located inside the Gum Nebula, given its estimated
distance of $d\sim 500$ pc. In this case, we do expect an enhanced DM
value without a similar enhancement in $N_H$. In fact, other known
pulsars at the same distance but located in different directions have
an average DM that is $\sim5$ times
smaller\footnote{http://www.atnf.csiro.au/research/pulsar/psrcat/}
than that of the \dpsr. Furthermore, all analyses of the X-ray
spectrum of the cooling neutron star RX J0720.4$-$3125 found $N_H$ in
the range $6-10\times10^{19}$ cm$^{-2}.$ Since RX~J0720.4$-$3125 and
the \dpsr\, are fortuitously close in position in the sky (the angular
distance is $\sim 4$ degrees) and at similar distances (360 pc for
RX~J0720.4$-$3125, Kaplan, van Kerkwijk \& Anderson
2007\nocite{kka07}), it seems reasonable to expect that their
foreground hydrogen columns should be similar.  We also note that the
$N_H$ resulting from fitting the X-ray spectrum of the \dpsr\, with
the explored non-thermal models corresponds to $\sim 30\%$ of the full
Galactic value obtained from the measurements of neutral hydrogen
(Dickey \& Lockman 1990\nocite{dl90}). On the other hand, the assumed
distance of \dpsr\, is $\sim 5\%$ of the neutral path length through
the disk along the line of sight to the system, as can be inferred
from e.g. Fig.~3a of McClure-Griffiths et
al. (2004\nocite{m-g+04}). These considerations would indicate a value
of $N_H$ for the \dpsr\, in the range $(0.2-0.3)\times10^{21}$
cm$^{-2},$ closer to the best fit value obtained for the thermal
models than for the non-thermal models. Taken at face value, this can
be interpreted as a weakness for the non-thermal emission
hypothesis. However, we also note that the value of $N_H$ resulting
from our fits with a PL (or a BB+PL) model is strongly dependent on
the adopted abundances: using the older abundances from Anders \&
Grevesse (1989\nocite{ag89}) we get a value of $N_H$ which is about
1/3 of that resulting from the more recent abundances (Lodders
2003\nocite{l03}) used in this paper (with all the other best fit
parameters almost unchanged).

The rotational energy loss $\dot{E}_A=5.8\times 10^{33}$
ergs s$^{-1}$ of pulsar A is similar to the values ($\dot{E}\sim
10^{33}-10^{34}$ ergs s$^{-1}$) of the fully recycled thermally
emitting MSPs (Zavlin 2006\nocite{z06}), whereas the fully recycled
non-thermally emitting MSPs have much larger $\dot{E}\sim
10^{35}-10^{36.5}$ ergs s$^{-1}.$ As a consequence, if a 
non-thermal emission is predominant in the \dpsr, the non-thermal
luminosity cannot be simply dependent on
the magnitude of the energy input from the pulsar.  Although the
surface magnetic fields $B_s$ of the thermally emitting MSPs are
comparable with those of the non-thermally emitting MSPs (with the
notable exception of PSR B1821$-$24), the values of their magnetic
field at the light cylinder $B_{lc}=B_s[2\pi R_{\rm ns}/cP]^3$ (where
$R_{\rm ns}$ is the neutron star radius and $P$ the spin period) are
different: $B_{lc}\sim 2-3\times 10^4$ G for the thermal emitters,
$B_{lc}\sim 2-9\times 10^5$ G for the non-thermal ones.  This may be
an ingredient for explaining the different emission properties of the
two classes of sources (Saito et al. 1997\nocite{skk+97}; Zavlin
2006\nocite{z06}), but the spectrum of pulsar A (in the hyphotesis
that it is non-thermal) does not fit this picture, since $B_{lc}\sim
5\times 10^3$ G, even lower than that of the MSPs with a thermal
spectrum.

What certainly differentiates pulsar A both from the thermally and the
non-thermally emitting MSPs (except PSR B1821$-$24, having
$B_s\sim 2\times 10^9$ G) is the much larger value of the magnetic
field close to the neutron star surface.  This could play a
significant role in a polar cap scenario. In particular, as
already noted by Chatterjee et al. (2007)\nocite{cgm+07}, pulsar A is one
of the very few recycled pulsars (see Fig.~1 of Harding, Usov \&
Muslimov 2005\nocite{hum05}) located above the death line for
curvature radiation of Harding \& Muslimov (2002\nocite{hm02};
PSR B1821$-$24 also satisfies this condition).  Detailed
calculations will be necessary to investigate if this property may
help trigger significant non-thermal X-ray emission from pulsar A,
despite the pulsar having a longer spin period and a smaller
$\dot{E}$ than the other non-thermal MSPs. The different values of
$P$ and $B_s$ with respect to the known population of X-ray
emitting MSPs may also be a factor for determining the unusually
steep photon index of pulsar A. In fact, all of the catalogued fully
recycled MSPs dominated by power-law emission show $\Gamma\lapp 2$
(although values of $\Gamma\sim 2.5-3.0$ are not unheard in the
population of non-recycled pulsars; see e.g. the recent
compilation of Li et al. 2007\nocite{lll07}).

Interestingly, the spectrum of the only other DNS detected in X-rays,
the B1534+12 system, can also be fitted with a
power-law\footnote{Given the poor photon statistics, at the moment the
data for PSR B1534+12 cannot exclude a blackbody model (Kargaltsev et
al. 2006\nocite{kpg06})} having a very soft photon index
$\Gamma=3.2\pm0.5$ (Kargaltsev et al. 2006\nocite{kpg06}), comparable
with that of the \dpsr.  Longer observations will be necessary to
better constrain this soft supposedly non-thermal spectrum and to
reveal if the X-ray flux is modulated at the spin period ($P=37.9$ ms)
of PSR B1534+12. A positive result would confirm the indication
emerging from our analysis, i.e. that a peculiarly soft X-ray emission
originates from the mildly recycled pulsars.

\subsection{Non-thermal emission: other processes}
\label{nonthermal:others} 

Two facts challenge the interpretation of the X-ray emission from the
\dpsr\, as due to a shock at the interface between pulsar A's wind and
pulsar B's magnetosphere (see the hypothesis (4a) described in
\S\ref{intro}). On one hand, we observe an apparent lack of modulation
of the X-ray flux with orbital phase (\S\ref{timing}), while on the
other hand, one detects a high degree of modulation at the spin-period
of A (Chatterjee et al.
2007\nocite{cgm+07}).  In fact, relativistic beaming of the X-ray
photons emitted at the shock front is expected to produce an orbital
modulation of order $\lapp 50\%$ (Arons \& Tavani 1993\nocite{at93},
Granot \& M\'esz\'aros 2004\nocite{gm04}, Pellizzoni et
al. 2004\nocite{pdm+04}). Despite the improved statistics resulting
from the combination of all the suitable photons collected so far, no
orbital variation of the X-ray flux from the \dpsr\, has been
detected, with an upper limit to the orbital modulation of $\sim 20\%$
(for an assumed sinusoidal variation). A further prediction of the
model (Granot \& M\'esz\'aros 2004\nocite{gm04}) is that variability
of the shock emission at the spin rate of pulsar A would be
significantly washed out due to the large ratio between the time of
flight of the X-ray photons from pulsar A to pulsar B's magnetosphere
and the rotational period of pulsar A. This contrasts with the very
high pulsed fraction of the light-curve folded at the period of pulsar
A.  As noticed by Chatterjee et al. (2007)\nocite{cgm+07}, the absence
of this kind of shock emission may support a very high degree of
magnetization of the pulsar wind close to pulsar A. In fact, for
highly magnetized shocks, the higher the degree of magnetization, the
smaller we expect the X-ray luminosity of the shock to be (Kennel \&
Coroniti 1984\nocite{kc84}). On the other hand, since the energy
budget is still favorable (up to about $3\times 10^{31}$ ergs s$^{-1}$
may in principle be available for powering this shock emission;
Lyutikov 2004\nocite{l04}), we cannot exclude that a component of the
observed total luminosity comes from this process, producing an
orbital modulation below our upper limit. In fact, a \CXO\,
observation of PSR B1534+12 indicates a deficit of X-ray emission
around apastron (Kargaltsev et al. 2006\nocite{kpg06}). Since the
orbit of the B1534+12 system is more eccentric ($e=0.274$) than that
of the \dpsr\, the shock emission may be much more modulated in the
former binary\footnote{As an alternate option, Kargaltsev et
al. (2006\nocite{kpg06}) suggest that the orbital X-ray modulation
seen in PSR B1534+12 is due to the misalignment between the equatorial
plane of the recycled pulsar and the binary orbit. This misalignment
may be larger for the B1534+12 system than for the \dpsr.}, while the
collision between pulsar A's wind and pulsar B's magnetosphere may still
provide a fraction of the total luminosity with a low level of
modulation. In particular, if the entire unpulsed emission ($\sim
25\%$ of the total luminosity of the \dpsr; Chatterjee et
al. 2007\nocite{cgm+07}), should be ascribed to this process, the
expected orbital modulation would be $\lapp 12\%.$

Energetic considerations make questionable an interpretation of the
X-ray photons as due to the emission of pulsar A's wind behind the
shock caused by the motion of the \dpsr\, in the interstellar medium
(hypothesis (4b) described in \S\ref{intro}). In particular, the
predicted X-ray flux for this process (Granot \& M\'esz\'aros
2004\nocite{gm04}) is $\sim 2\times 10^{29}$ ergs s$^{-1}$ (assuming a
typical particle density of 10 cm$^{-3}$ in the interstellar medium
and the observed systemic velocity of the \dpsr, $\sim 10$ km
s$^{-1}$; Kramer et al. 2006\nocite{ksm+06}), i.e. an order of
magnitude less than that observed (see Table~1).  We note that the
derived upper limits ($\sim 2\times 10^{30}$ ergs s$^{-1}$, see
\S\ref{spect}) on the luminosity of any diffuse emission around the
position of the \dpsr\, are not yet very constraining for the presence
of a PWN.  In fact, adopting the recent scaling law of Li et
al. (2007)\nocite{lll07} (and allowing for the typical scatter of one
order of magnitude in the correlation) the expected luminosity in the
0.3--8 keV band of a PWN (if any) powered by the rotational energy
loss of pulsar A ($\dot{E}_A=5.8\times 10^{33}$ ergs s$^{-1}$) should
be $\lapp 4\times 10^{29}$ ergs s$^{-1}.$ Nevertheless, this
luminosity estimate, the high pulsed fraction of the emission at the
spin period of pulsar A and the very soft spectrum (strongly at
variance with the photon indexes seen in PWNe, see \S\ref{intro})
permit us to conclude that a putative PWN can give only a negligible
contribution to the energy budget of the X-ray emission from the
\dpsr.

\subsection{Summary}

Our analysis unambiguously indicates that the X-ray emission of the
\dpsr\, is characterized by a very soft spectrum.  However, on the
basis of the available data, we cannot discriminate between a
predominantly thermal and a predominantly non-thermal origin for this
spectrum.

In the thermal emission hypothesis, the X-ray photons of the \dpsr\,
may (i) originate from two roughly concentric regions on the surface of
pulsar A, heated to different temperatures by particles accelerated in
the magnetosphere of pulsar A and impinging onto the neutron star, or
may (ii) result from the combination of heated polar cap(s) emission and
non-thermal photons scattered in inverse Compton processes in the
pulsar magnetosphere. The high pulsed fraction (at the spin period
of pulsar A) of the X-ray emission is a problem for these
interpretations.

In the predominantly non-thermal emission hypothesis, magnetospheric
emission from pulsar A may provide the bulk of the X-ray flux
and easily explain the high pulsed fraction of the emission. A
smaller contribution to the X-ray luminosity - due to either the
interaction between pulsar A's wind and pulsar B's magnetosphere or to
a thermal emission from heated caps - cannot be excluded, but it is
not required by the available observations. The relatively high value of
$N_H$ (when compared with the assumed distance of the source) is
the main difficulty with this interpretation.

The very soft nature of the spectrum appears to be echoed in the
spectrum of the only other mildly recycled pulsar detected so far in
the X-ray band, PSR B1534+12. This may corroborate the hypothesis
that mildly recycled pulsars constitute a new class of neutron star
X-ray emitters, undergoing different emission processes with respect
to those modeled for the fully recycled pulsars. If this holds true,
their study will be particularly important for understanding how
different values of spin period and surface magnetic field may affect
the mechanisms of X-ray production in rotation-powered neutron
stars.

\acknowledgements {\small{The authors thank the anonymous referee for
a careful reading of the manuscript, and the very helpful comments and
suggestions. AP and MB acknowledge the financial support to this
research provided by the {\it Ministero dell'Istruzione,
dell'Universit\`a e della Ricerca} (MIUR) under the national program
{\it PRIN05 2005024090\_002}. NR is supported by an NWO Veni
Fellowship, and MAM acknowledges support from WVEPSCoR. 
This work was supported in part by SAO grant G05-6044.}}


\begin{thebibliography}{44}
\expandafter\ifx\csname natexlab\endcsname\relax\def\natexlab#1{#1}\fi
\bibitem[Anders \& Grevesse]{ag89}
Anders, E., \& Grevesse, N. 1989, Geochim. Cosmochim. Acta, 53, 197
\bibitem[Arons (1981)]{a81}
Arons, J. 1981, ApJ, 248, 1099
\bibitem[Arons \& Tavani (1993)]{at93}
Arons, J., \& Tavani, M. 1993, ApJ, 403, 249
\bibitem[Bogdanov, Grindlay \& Rybicki (2006)]{bgr06}
Bogdanov, S., Grindlay, J.E., \& Rybicki, G.B. 2006, ApJ, 648, L55
\bibitem[Burgay et~al. (2003)]{bdp+03}
Burgay, M., et al. 2003, Nature, 426, 531
\bibitem[Campana et~al.(2004)]{cpb04}
Campana, S., Possenti, A., \& Burgay, M. 2004, ApJ, 613, L53
\bibitem[Chatterjee, Goss \& Brisken (2005)]{cgb05}
Chatterjee, S., Goss, W. M., \& Brisken, W.F. 2005, ApJ, 634, 101
\bibitem[Chatterjee et al. (2007)]{cgm+07}
Chatterjee, S., Gaensler, B.M., Melatos, A. Brisken, W.F., \&
Stappers, B.W. 2007, ApJ, 670, 1301
\bibitem[Cheng \& Ruderman (1980)]{cr80}
Cheng, A.F., \& Ruderman, M.A. 1980, ApJ, 235, 576
\bibitem[Cheng \& Zhang (1999)]{cz99}
Cheng, K.S., \& Zhang, L. 1999, ApJ, 515, 337
\bibitem[Cheng. Taam \& Wang (2006)]{ctw06}
Cheng, K.S., Taam, R.E., \& Wang, W. 2006, ApJ, 641, 427
\bibitem[Cordes \& Lazio (2002)]{cl02}
Cordes, J.M.,\& Lazio, T.J.W., 2002, preprint (astro-ph/0207156)
\bibitem[de Jager, Swanepoel \& Raubenheimer (1989)]{dsr89}
de Jager, O.C., Swanepoel, J. W.H., \& Raubenheimer, B.C. 1989, A\&A, 221, 180
\bibitem[Dickey \& Lockman (1990)]{dl90}
Dickey, J.M., \& Lockman, F.J. 1990, ARA\&A, 28, 215
\bibitem[Gaensler \& Slane (2006)]{gs06}
Gaensler, B.M., \& Slane, P.O. 2006, ARA\&A, 44, 17
\bibitem[Gehrels (1986)]{g86}
Gehrels, N. 1986, ApJ, 303, 336
\bibitem[Geppert, K\"uker \& Page (2006)]{gkp06}
Geppert, U., K\"uker, M., Page, D. 2006, A\&A, 457, 937
\bibitem[Gonzalez et al. (2005)]{gkc05} Gonzalez, M.~E., Kaspi, V.~M.,
Camilo, F., Gaensler, B.~M., Pivovaroff, M.~J. 2005, ApJ, 630, 489
\bibitem[Granot \& M\'esz\'aros (2004)]{gm04}
Granot, J., \& M\'esz\'aros, P. 2004, ApJ, 609, L17
\bibitem[Grindlay et al. (2002)]{gch+02}
Grindlay, J.E., Camilo, F., Heinke, C.O., Edmonds, P.D., Cohn, H., \&
Lugger, P. 2002, ApJ, 581, 470
\bibitem[Harding, Usov \& Muslimov (2005)]{hum05}
Harding, A.K., Usov, V.V., \& Muslimov, A.G. 2005, ApJ, 622, 531
\bibitem[Harding, \& Muslimov (2002)]{hm02}
Harding, A.K., \& Muslimov, A.G. 2002, ApJ, 568, 862
\bibitem[Kaplan, van Kerkwijk \& Anderson (2007)]{kka07}
Kaplan,D.L., van Kerkwijk, M.H., \& Anderson, J.  2007, ApJ, 660, 1428 
\bibitem[Kargaltsev, Pavlov \& Garmire (2006)]{kpg06}
Kargaltsev, 0., Pavlov, G.G.,  \& Garmire G.P. 2006, ApJ, 646, 1139
\bibitem[Kennel \& Coroniti 1984]{kc84}
Kennel, C.F., \& Coroniti, F.V. 1984, ApJ, 283, 694
\bibitem[Kramer et al. (2006)]{ksm+06}
Kramer, M., et al. 2006, Science, 314, 97
\bibitem[Li, Lu \& Li (2007)]{lll07}
Li, X-H., Lu, F-J. \& Li, Z. 2007, ApJ, submitted (arXiv:0707.4279)
\bibitem[Lodders (2003)]{l03}
Lodders, K. 2003, ApJ, 591, 1220
\bibitem[Lorimer et al. (2007)]{lfs+07}
Lorimer, D. R., et al. 2007, MNRAS, 379, 1217
\bibitem[Lyne et al. (2004)]{lbk+04}
Lyne, A. G., et al. 2004, Science, 303, 1153
\bibitem[Lyutikov (2004)]{l04}
Lyutikov, M. 2004, MNRAS, 353, 1095
\bibitem[McLaughlin et~al.(2004a)]{mkl+04}
McLaughlin, M.~A., et al. 2004a, ApJ, 613, L57
\bibitem[McLaughlin et~al. (2004b)]{mll+04}
McLaughlin, M.~A., et al. 2004b, ApJ, 616, L131
\bibitem[McLaughlin et al.(2004c)]{mcb+04}
McLaughlin, M. A., et al. 2004c, ApJ, 605, L41
\bibitem[McClure-Griffiths et al. (2004)]{m-g+04} 
McClure-Griffiths, N.~M., Dickey, J.~M., Gaensler, B.~M., \& Green,
A.~J. 2004, ApJ, 607, 127
\bibitem[Page \& Sarmiento (1996)]{ps96}
Page, D., \& Sarmiento, A. 1996, ApJ, 473, 1067
\bibitem[Pellizzoni et~al.(2004)]{pdm+04}
Pellizzoni, A., De Luca, A., Mereghetti, S., Tiengo, A., Mattana, F.,
Caraveo, P., Tavani, M., \& Bignami, G.F. 2004, ApJ, 612, L49
\bibitem[Pellizzoni et~al.(2008)]{p+08}
Pellizzoni, A., Tiengo, A., De Luca, A., Esposito, P., \& Mereghetti, S. 2008, 
ApJ, in press, (arXiv:0802.0350)
\bibitem[Possenti et al. (2002)]{pcc+02}
Possenti, A., Cerutti, R., Colpi, M., \& Mereghetti, S. 2002, A\&A, 387, 993
\bibitem[Psaltis, \"Ozer \& DeDeo (2000)]{pod00}
Psaltis, D., Özel, F., \& DeDeo, S. 2002, ApJ, 544, 390
\bibitem[Saito et al. (1997)]{skk+97}
Saito, Y., Kawai, N., Kamae, T., Shibata, S., Dotani, T., \& Kulkarni, S.R.
1997, ApJ, 447, L37
\bibitem[Stairs et al.(2002)]{stt+02}
Stairs, I.H., Thorsett, S.E., Taylor, J.H., \& Wolszczan, A. 2002,
ApJ, 581, 501
\bibitem[Turolla \& Treves 2004]{tt04}
Turolla, R., \& Treves, A. 2004, A\&A, 426, L1
\bibitem[Yakovlev \& Pethick (2004)]{yp04}
Yakovlev, D. G., \& Pethick, C. J., 2004, Ann. Rev. A\&A, 42, Issue 1, 169
\bibitem[Zhang \& Harding (2000)]{zh00}
Zhang, B., \& Harding, A.K. 2000, ApJ, 532, 1150
\bibitem[Zavlin, Pavlov \& Shibanov (1996)]{zps96}
Zavlin, V.E.,Pavlov G.G, \& Shibanov, Yu.A. 1996, A\&A, 315, 141
\bibitem[Zavlin \& Pavlov (1998)]{zp98}
Zavlin, V.E., \& Pavlov, G.G, 1998, A\&A, 329, 583
\bibitem[Zavlin (2006)]{z06}
Zavlin, V.E. 2006, ApJ, 638, 951
\end{thebibliography}

\begin{table}
\footnotesize{
\begin{center}
\caption{Observations of the PSR~J0737--3039 system}
\vspace{0.3cm}
\tabcolsep=0.06truecm
\begin{tabular}{lccccccc}
\hline
\hline
Instrument    & Start date & Tot exp$^a$ & Tot Source (Bkg)$^b$ & Orbits$^c$  & Source (Bkg)$^d$ & Analysis$^e$ & Ref$^f$ \\
              & & (ks)        &   (counts) &             &   (counts)       &              &         \\
\hline
Chandra ACIS-S    & 2004/01/18 & 10.0 & 70    (1) & 1   &  64   (1) & t,p   & (1),(4) \\
XMM MOS1+MOS2$^g$ & 2004/04/10 & 47.2 & 549 (155) & 5+4 & 432 (110) & t,s   & (2),(3),(4)\\
Chandra HRC-S     & 2006/02/28 & 53.5 & 251   (5) & 6   & 250   (5) & t,p   & (5) \\
Chandra HRC-S     & 2006/03/02 & 35.8 & 169   (6) & 4   & 167   (6) & t,p   & (5) \\
Chandra ACIS-S    & 2006/06/06 & 79.0 & 534   (3) & 9   & 529   (3) & t,p,s & this work \\
\hline
\tablecomments{Observations used for spectral, spatial and timing
analysis. \\
$^a$ Total exposure times calculated after cleaning
for spurious flares (for \XMM), but not including dead-time
corrections.\\
$^b$ The source (and background: Bkg) counts are reported in the bands
used for the timing analysis: 0.3--8 keV (\CXO) and 0.3--2.2 keV
(\XMM). The background dominates above 2.2 keV in the \XMM\, band
(Pellizzoni et al. 2004\nocite{pdm+04}). The time resolutions are 16
$\mu$s for HRC-S, 2.6 s for MOS1+MOS2, and 3.241 s for ACIS-S.\\
$^c$ Number of sampled full orbits.\\
$^d$ The source (and background: Bkg) counts refer to the number of
completely sampled orbits of column 5.\\
$^e$ Analysis which a dataset has been used for in this work:
 p=spatial; s=spectral; t=timing.\\
$^f$ References: (1) McLaughlin et al. 2004c\nocite{mcb+04}; (2)
Campana et al.~2004\nocite{cpb04}; (3) Pellizzoni et
al.~2004\nocite{pdm+04}; (4) Kargaltsev et al.~2006\nocite{kpg06};
(5) Chatterjee et al.~2007\nocite{cgm+07}. \\
$^g$ The photons collected with the PN detector (operated in timing
mode) have not been included in our analysis, since they are largely
background dominated (Pellizzoni et al. 2004\nocite{pdm+04}).}
\end{tabular}
\end{center} }
\label{tab:1}
\end{table}

\begin{table}
\footnotesize{
\begin{center}
\caption{Spectral parameters for the PSR J0737--3039 system}
\vspace{0.3cm}
\tabcolsep=0.08truecm
\begin{tabular}{lcccccc}
\hline
\hline
Model & BB & PL & Bremss & BB+PL  & BB+BB  &Comptonized BB\\
\hline
$N_{H}$$^a$                 & $<0.1$ &
1.6$\pm$0.6                 & $<0.1$           & 1.5$\pm$0.8  & $<0.1$  & $<0.1$ \\
$\Gamma$/$kT_1$/$kT_2$$^b$  & 0.18$\pm$0.05 &
3.7$\pm$0.4                 & 0.6$\pm$0.1      & 3.7$\pm$0.5/0.18$^e$       &
0.10$\pm$0.01/0.30$\pm$0.05 & 0.07$\pm$0.03$^f$ \\
Flux$^c$                    & 2.5$\pm$1 &
8$\pm$1                     & 3$\pm$1          & 8$\pm$1      & 5$\pm$1     & 5$\pm$1\\
Luminosity$^d$              & 0.7$\pm$0.3 &
2.4$\pm$0.3                 & 0.9$\pm$0.3      & 2.4$\pm$0.3  & 1.5$\pm$0.3 & 1.5$\pm$0.3\\
\hline
$\chi^2_\nu$;~d.o.f.;~n.h.p.& 2.32;~47;~$10^{-7}$ & 1.25;~47;~0.09             
& 1.27;~47;~0.08 & 1.29;~46;~0.07 & 1.23;~45;~0.08 & 1.25;~45;~0.18\\
\hline
\end{tabular}
\tablecomments{Parameter values of the spectral models discussed in the text. 
Errors are reported at 90\% confidence level. \\
$^a$ $N_{H}$ is in units of $10^{21}$cm$^{-2}$ assuming solar
abundances from Lodders (2003)\nocite{l03}. \\
$^b$ $kT_1$ and $kT_2$ are in keV.\\
$^c$ Fluxes are unabsorbed, calculated in the 0.3--8\,keV energy
range, and reported in units of $10^{-14}$~ergs\,s$^{-1}$\,cm$^{-2}$. \\
$^d$ Luminosities are expressed in units of $10^{30}$~ergs\,s$^{-1}$
and are calculated for isotropic emission at a distance of 500 pc. \\
$^e$ This temperature corresponds to the smallest $\chi^2$ value in a
pure blackbody fit, even though the fit is not acceptable.  Here it was
held fixed at this value. \\
$^f$ The value of the optical depth for the best fit is $\tau=$1.5$\pm$0.6
for a thermal electron bath with energy $kT_e=$10$\pm$5 keV}
\end{center} }
\label{tab:2}
\end{table}

\end{document}